\documentclass[onecolumn]{aastex61}
\newcommand{\target}{`Oumuamua}
\newcommand{\fulltarget}{1I/2017 U1~(\target)}
\newcommand{\targcaps}{`OUMUAMUA}

\begin{document}

\title{A Serendipitous MWA Search for Narrow-band and Broad-band Low Frequency Radio Transmissions from 1I/2017 U1 \targcaps}

\author[0000-0002-8195-7562]{Tingay, S.J.}
\affiliation{International Centre for Radio Astronomy Research, Curtin University, Bentley, WA 6102, Australia}

\author[0000-0001-6295-2881]{Kaplan, D.~L.}
\affiliation{Department of Physics, University of Wisconsin--Milwaukee, Milwaukee, WI 53201, USA}

\author[0000-0002-9994-1593]{Lenc, E.}
\affiliation{Sydney Institute for Astronomy, School of Physics, The University of Sydney, NSW 2006, Australia}
\affiliation{ARC Centre of Excellence for All-sky Astrophysics (CAASTRO)}

\author[0000-0003-4823-129X]{Croft, S.}
\affiliation{Department of Astronomy, University of California Berkeley, Berkeley, CA 94720, USA}

\author{McKinley, B.}
\affiliation{International Centre for Radio Astronomy Research, Curtin University, Bentley, WA 6102, Australia}

\author{Beardsley, A.}
\affiliation{School of Earth and Space Exploration, Arizona State University, Tempe, AZ 85287, USA}

\author{Crosse, B.}
\affiliation{International Centre for Radio Astronomy Research, Curtin University, Bentley, WA 6102, Australia}

\author{Emrich, D.}
\affiliation{International Centre for Radio Astronomy Research, Curtin University, Bentley, WA 6102, Australia}

\author{Franzen, T.M.O.}
\affiliation{International Centre for Radio Astronomy Research, Curtin University, Bentley, WA 6102, Australia}

\author{Gaensler, B.M.}
\affiliation{ARC Centre of Excellence for All-sky Astrophysics (CAASTRO)}
\affiliation{Dunlap Institute for Astronomy and Astrophysics, University of Toronto, ON, M5S 3H4, Canada}

\author{Horsley, L.}
\affiliation{International Centre for Radio Astronomy Research, Curtin University, Bentley, WA 6102, Australia}

\author{Johnston-Hollitt, M.}
\affiliation{International Centre for Radio Astronomy Research, Curtin University, Bentley, WA 6102, Australia}

\author{Kenney, D.}
\affiliation{International Centre for Radio Astronomy Research, Curtin University, Bentley, WA 6102, Australia}

\author{Morales, M.F.}
\affiliation{Department of Physics, University of Washington, Seattle, WA 98195, USA}

\author{Pallot, D.}
\affiliation{International Centre for Radio Astronomy Research, University of Western Australia, Crawley 6009, Australia}

\author{Steele, K.}
\affiliation{International Centre for Radio Astronomy Research, Curtin University, Bentley, WA 6102, Australia}

\author{Trott, C.M.}
\affiliation{International Centre for Radio Astronomy Research, Curtin University, Bentley, WA 6102, Australia}
\affiliation{ARC Centre of Excellence for All-sky Astrophysics (CAASTRO)}

\author{Walker, M.}
\affiliation{International Centre for Radio Astronomy Research, Curtin University, Bentley, WA 6102, Australia}

\author{Wayth, R.B.}
\affiliation{International Centre for Radio Astronomy Research, Curtin University, Bentley, WA 6102, Australia}
\affiliation{ARC Centre of Excellence for All-sky Astrophysics (CAASTRO)}

\author{Williams, A.}
\affiliation{International Centre for Radio Astronomy Research, Curtin University, Bentley, WA 6102, Australia}

\author{Wu, C.}
\affiliation{International Centre for Radio Astronomy Research, University of Western Australia, Crawley 6009, Australia}

\begin{abstract}
We examine data from the Murchison Widefield Array (MWA) in the frequency range 72 -- 102 MHz for a field-of-view  that serendipitously contained the interstellar object \target\ on 2017 November 28. Observations took place with time resolution of 0.5\,s and frequency resolution of 10\,kHz.  
Based on the interesting but highly unlikely suggestion that \target\ is an interstellar spacecraft, due to some unusual orbital and morphological characteristics, we examine our data for signals that might indicate the presence of intelligent life associated with \target.  We searched our radio data for: 1) impulsive narrow-band signals; 2) persistent narrow-band signals; and 3) impulsive broadband signals.  We found no such signals with non-terrestrial origins and make estimates of the upper limits on Equivalent Isotropic Radiated Power (EIRP) for these three cases of approximately 7\,kW, 840\,W, and 100\,kW, respectively.  These transmitter powers are well within the capabilities of human technologies, and are therefore plausible for alien civilizations.  While the chances of positive detection in any given Search for Extraterrestrial Intelligence (SETI) experiment are vanishingly small, the characteristics of new generation telescopes such as the MWA (and in the future, the Square Kilometre Array) make certain classes of SETI experiment easy, or even a trivial by-product of astrophysical observations. This means that the future costs of SETI experiments are very low, allowing large target lists to partially balance the low probability of a positive detection.
\end{abstract}


\section{Introduction}

The recently-discovered object \fulltarget\ is apparently visiting our solar system on a hyperbolic trajectory \citep{pmid29160305}. It lacks a cometary coma, and its shape appears to be elongated when compared to solar system asteroids. However, the presence of such objects passing through our solar system is not unexpected. Interstellar dust grains have been detected in large numbers \citep[e.g.,][]{2016Sci...352..312A}, and theories of planet formation predict that much larger fragments may also be ejected into interstellar space \citep{2003Icar..166..141C}; some such objects would be expected to encounter other planetary systems as they roam the Galaxy.

The most likely explanation for the origin of \target\ is that it is a cometary fragment that has lost much of its surface water due to bombardment by cosmic rays during its long journey through interstellar space \citep{fitzsimmons2017spectroscopy}. However, if advanced civilizations exist elsewhere in our Galaxy, it is feasible to speculate that they may develop the capability to launch spacecraft over interstellar distances \citep{1960Natur.186..670B,1980JBIS...33...95F}, and that these spacecraft may use radio waves to communicate. The ratio of the number of such craft to the number of interstellar cometary fragments in our solar system is presumably low, if not zero, but we ought not to be so complacent as to rule out an artificial origin for objects such as \target\ without first performing some observations.

As the first object of its class to be discovered, \target\ provides an interesting opportunity to expand the search for extraterrestrial intelligence (SETI) from traditional targets such as stars and galaxies \citep[e.g.][]{2017PASP..129e4501I}, to objects that are much closer to Earth. This also allows for searches for transmitters that are many orders of magnitude fainter than those that would be detectable from a planet orbiting even the most nearby stars.

It is difficult to estimate the chance of success of programs that attempt to constrain the number of transmitting civilizations in our Galaxy, but it is clear that while the chances that any single target will reveal signatures of extraterrestrial technology are low, our chances of success increase as the number of targets surveyed increases. We therefore chose to undertake a SETI search of \target, the results of which (non-detections of the signals searched for) are reported here as a prototype for similar searches to be undertaken in future.

This study used serendipitous observations from the Murchison Widefield Array \citep[MWA;][]{2013PASA...30....7T}, the Square Kilometre Array low-frequency precursor. The MWA's wide field of view and extremely radio quiet site make it ideally suited to SETI work, and planned upgrades to the telescope's back-end systems will soon enable routine commensal searches for signatures of technology from a wide range of targets across the sky visible from the site.  This paper thus continues our exploration of the MWA's capabilities for SETI experiments.

\section{Observations and Data Processing}

\subsection{Identification of Observations}
We searched the MWA data archive for serendipitous MWA observations of \target\ over the period 2017~November~01 to 2018~January~10, corresponding to the period from the start of regular observing with the extended MWA Phase II array up to the present.  The range to \target\ varied from 0.63\,AU to 3.96\,AU over that period.  To search we used the JPL Horizons service\footnote{See \url{https://ssd.jpl.nasa.gov/horizons.cgi}} to return the apparent topocentric celestial coordinates of \target\ from the MWA site at 1 minute intervals.  We then compared these positions against all of the MWA observations during that time period and identified observations where the separation between the position of \target\ and the MWA's primary beam center was $<15\degr$.  We found a promising observation from project G0017 (PI: B.~McKinley. Pointing centre: RA=351.616$^{\circ}$; DEC=$-$2.63$^{\circ}$ (J2000)), summarized in Table~\ref{tab:obs}.  The observation used a bandwidth of 30.72\,MHz covering 72--102\,MHz (including part of the FM band), and recorded visibilities with time resolution 0.5\,s and frequency resolution 10\,kHz. 

\subsection{Calibration, Imaging, and Extraction of Dynamic Spectra}

The real-time calibration and imaging system \citep[\textsc{rts};][]{Mitchell:2008, Ord:2010} was used for bandpass and gain calibration.  Calibration was performed using a snapshot observation of Hydra A and a simple point-source model. Visibility data from baselines shorter than $50\lambda$ were down-weighted to improve calibration in the presence of diffuse structure, and baselines longer than $1000\lambda$ were excluded to avoid resolving the calibrator source. Archived flagging was applied to the calibrator data to remove radio frequency interference (RFI). We estimate the uncertainty on the absolute amplitude calibration to be better than 10\%.

\begin{figure}
\plotone{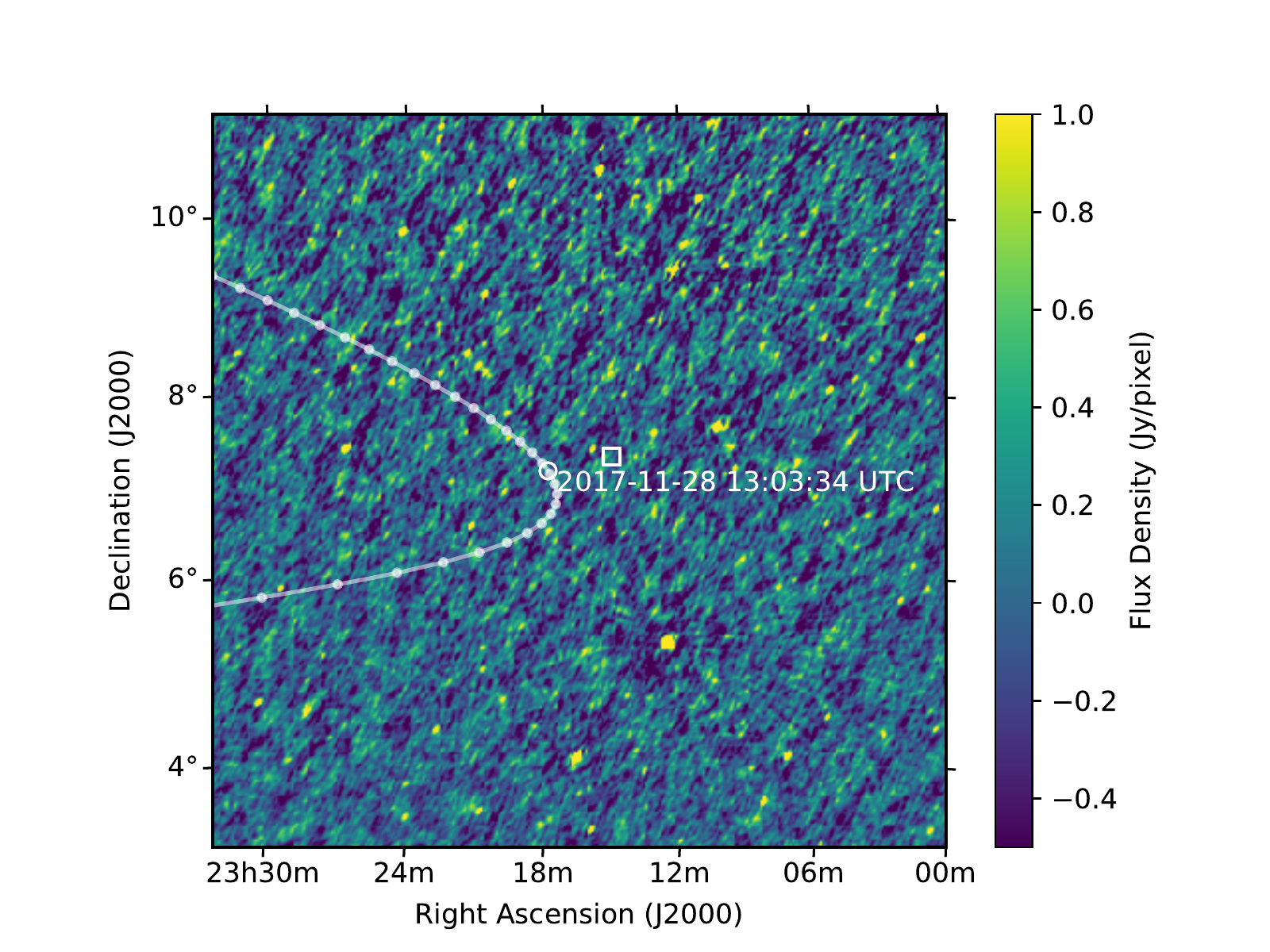}
\caption{Dirty continuum image of the field around \target.  The bright source PKS~J2316+0405, located just off the bottom of the image, has been peeled to remove it from the data. The path of \target\ is shown with the white line, where dots are plotted every two days.  The position of \target\ at the time of the  observation analyzed here is shown with the open circle.  The off-source comparison position is the open square.}
\label{fig:image}
\end{figure}

The calibration solutions from the Hydra A observation were applied to the target dataset and a dirty image cube was created using the \textsc{rts} at full 10\,kHz spectral resolution over a $8\arcdeg\times8\arcdeg$ region centered on the location of \target. Natural weighting was used to increase sensitivity and baselines shorter than $50\lambda$ were excluded to minimize sidelobe confusion and sensitivity to large-scale structure. No RFI flagging was performed on the target observations, except to remove 17 out of 128 fine (10\,kHz) channels from each of the 24 coarse (1.28\,MHz) channels affected by the poly-phase filter bank \citep{Ord:2015}, resulting in an image cube with $2664\times10$\,kHz channels. Channel images with an image noise exceeding 1.8 times the median noise were flagged (33 fine channels were found to be affected in this manner). A continuum image (see Figure~\ref{fig:image}) was created by averaging the spectral cube in frequency. The resulting continuum map was examined to ensure that the data were not corrupted by poor calibration or local RFI.

\begin{figure}
\plottwo{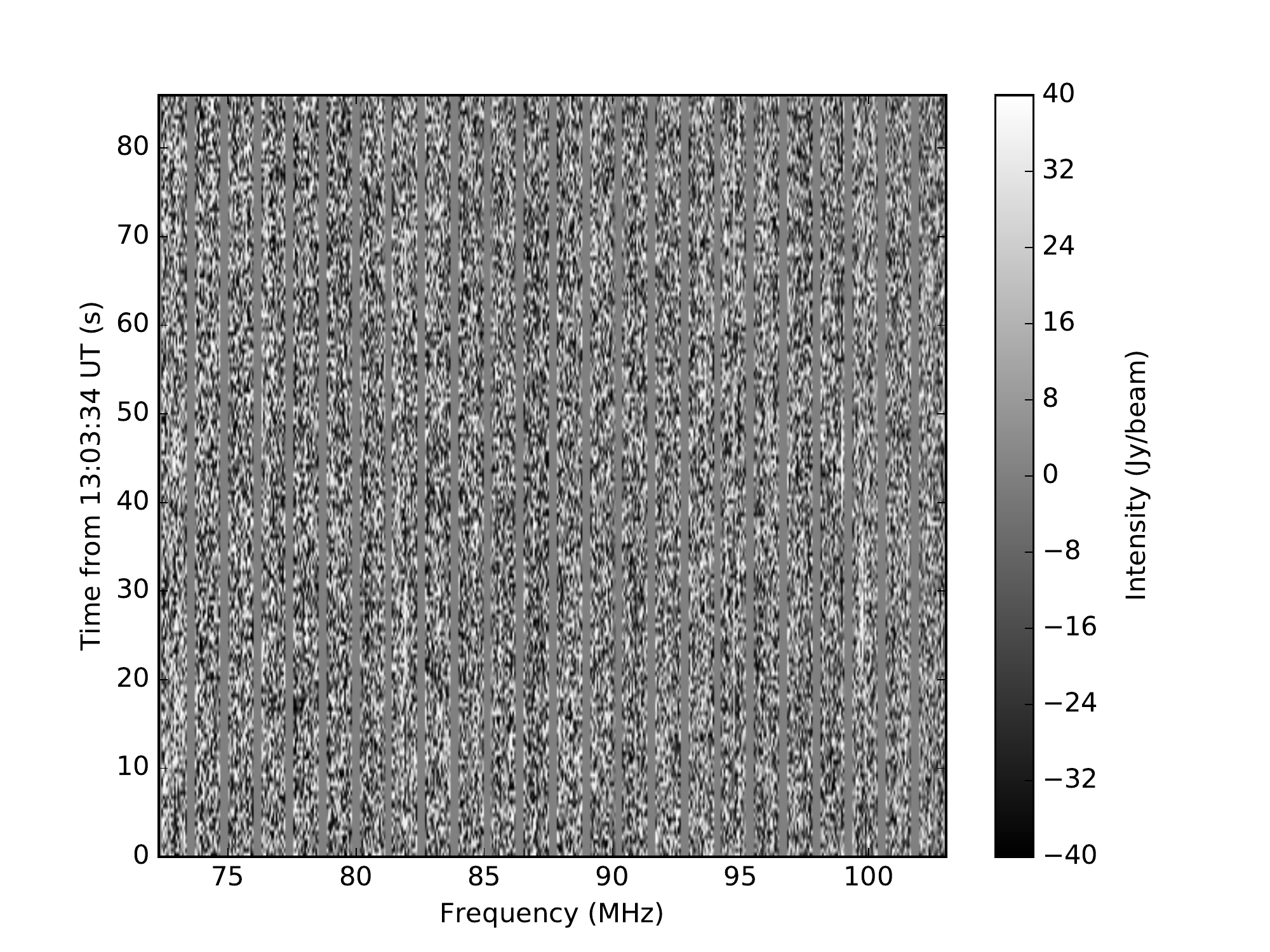}{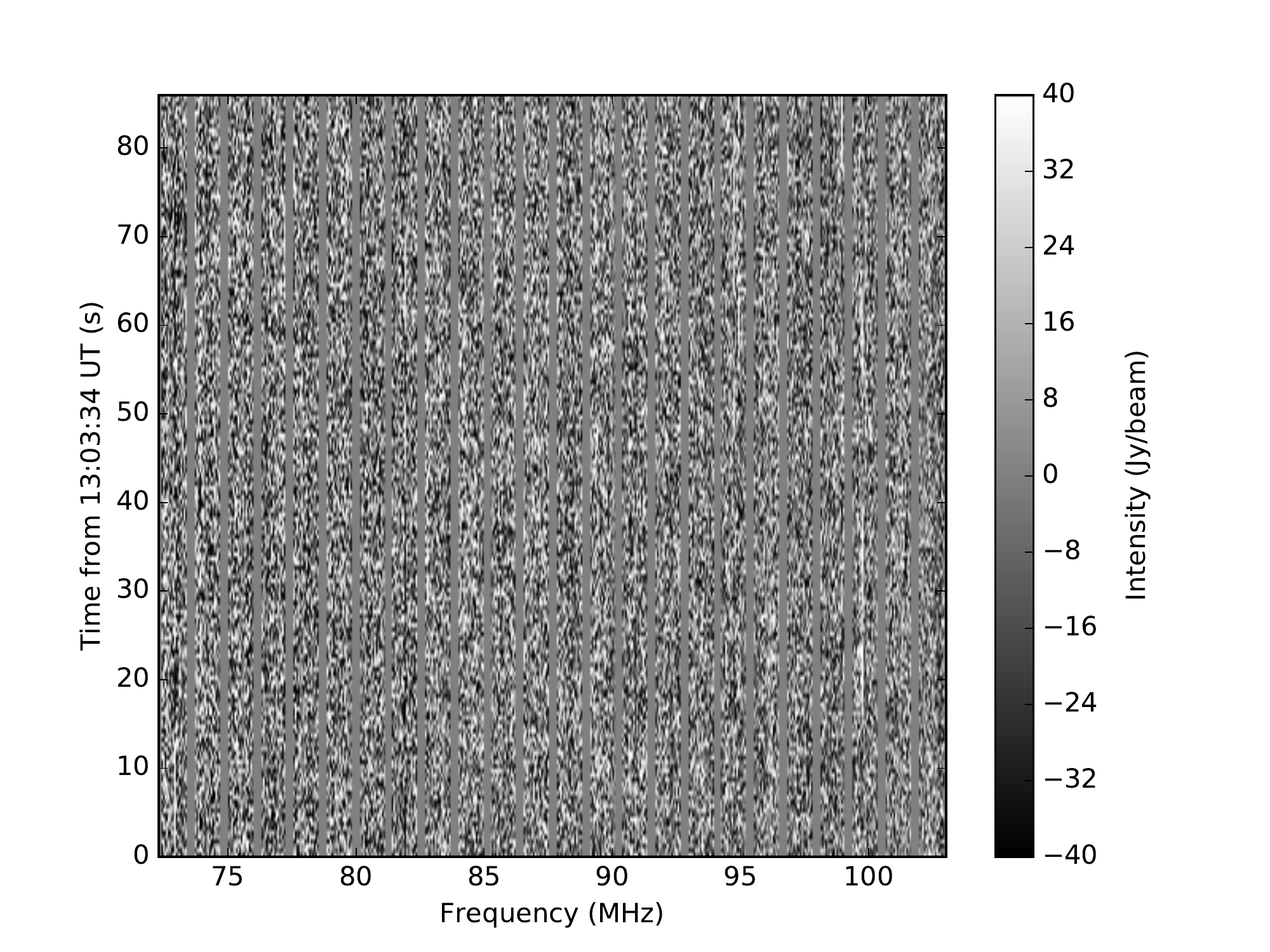}
\caption{Dynamic spectra at the on (left panel) and off (right panel) positions described in the text and indicated in Figure~\ref{fig:image}.  Sixteen channels at each edge of each coarse channel have been blanked (set to zero), to avoid artifacts caused by the coarse polyphase filter bank.  The time resolution is 0.5\,s, the frequency resolution is 10\,kHz and the amplitude scale is clipped at $\pm$40\,Jy in order to show the noise levels.}
\label{fig:ds}
\end{figure}

To allow extraction of dynamic spectra, the visibility data were re-imaged at full spectral resolution (10\,kHz) and an image cube was generated for each 0.5\,s integration over a $\sim$86\,s period. The size of the images was reduced to $2\arcdeg\times2\arcdeg$ to limit storage requirements.  The final images have an approximate 3\arcmin\ angular resolution.  Once we had the dirty image cubes, we extracted dynamic spectra at the position of \target\ and a comparison location ($23^{\rm h}15^{\rm m}20\fs72$, $+7\degr15\arcmin23\farcs0$, $0.7\degr$ away).  Figure~\ref{fig:ds} shows the dynamic spectra for both on and off positions.

\begin{deluxetable}{c c c c c c c c}
\tabletypesize{\footnotesize}
\tablewidth{0pt}
\tablecaption{MWA Observations of \fulltarget\label{tab:obs}}
\tablehead{
\colhead{Start} & \colhead{Duration} & \colhead{Frequency} & \multicolumn{2}{c}{Position of \target\tablenotemark{a}} & \colhead{Separation\tablenotemark{b}} & \colhead{Primary Beam\tablenotemark{c}} & \colhead{Range}\\
\colhead{(UT)} & \colhead{(s)} & \colhead{(MHz)} & \colhead{RA} & \colhead{Dec} & \colhead{(deg)} & \colhead{(\%)} & \colhead{(AU)}
}
\startdata
2017-11-28 13:03:34 & 86 & 72--102 & $23^{\rm h}18^{\rm m}08\fs33$ & $+07\degr05\arcmin56\farcs3$ & \phn9.9 & 64 &1.68\\
\enddata
\tablenotetext{a}{Apparent position from the MWA site (J2000).}
\tablenotetext{b}{Separation between pointing boresight and position of \target.}
\tablenotetext{c}{Primary beam response compared to boresight.}
\end{deluxetable}

\section{Results}

The dynamic spectra were examined to search for signals of the following types, isolated to the position of \target: 1) narrow-band impulsive signals i.e., within a single 10\,kHz and 0.5\,s pixel of the dynamic spectrum; 2) persistent narrow-band signals over the duration of the observation; and 3) impulsive broad-band signals (i.e., within single 0.5\,s time steps).  Persistent broad-band signals were not searched for, since differences in total power between different locations in the field are dominated by the locations of background radio sources (including a background of confused sources).  Signals of the three types listed above use some level of isolation in time and/or frequency to overcome this issue.  In all tested cases, the off position was utilised for comparison to the on position, when determining if a candidate signal was isolated to \target.

The use of on/off comparisons to isolate signals of various types on the sky is common practise using single dishes, assuming that the same terrestrial RFI signals are equally present at both on and off positions closely spaced on the sky, which is generally a good assumption.  The situation for an interferometer is more complicated than for a single dish.  The visibility response of an interferometer to RFI depends on the position of the interferer (generally very far from the visibility phase centre and in the near field of the interferometer) relative to the various baseline orientations and lengths present in the interferometer.  The response therefore varies strongly as a function of baseline and time.  The Fourier transform relationship between visibilities and image plane transfers this complexity to the image plane, meaning that for the on and off position pixels in our images, their values as a function of time will not be the same.  However, while not the same, both pixels will experience large deviations from noise-like behaviour and this is the signature that will distinguish terrestrial RFI from signals confined to \target\ (in the far field).


\subsection{Narrow-band Impulsive Signals}
Using the dynamic spectrum for the on position (Figure~\ref{fig:image}), we measured the root mean square (RMS) variation of the pixels to be approximately 50\,Jy/beam.  While the majority of the signal in the dynamic spectrum is noise-like, the presence of RFI means that the signal contains a high amplitude tail compared to a Gaussian distribution.  Strictly, Gaussian statistics cannot be used to calculate detection probabilities in this case.  For example, with our dataset size, the assumption of Gaussian statistics gives the expectation of approximately 0.1 events above 5$\sigma$, whereas at greater than five times the measured RMS in the dynamic spectrum, we find 141 pixels. Given the close to Gaussian, but strictly unknown underlying probability density function of our data, we make a subjective choice to use a simple threshold of five times the measured RMS to identify candidate detections.  The 141 pixels above this threshold at the on position are contained in 25 individual 10\,kHz frequency channels.  These frequency channels are listed in Table~\ref{tab:chans}.

\begin{deluxetable}{c c c}
\tabletypesize{\footnotesize}
\tablewidth{0pt}
\tablecaption{Frequency channels affected by RFI at the on position\label{tab:chans}}
\tablehead{
\colhead{Channel \#} & \colhead{Frequency} & FM call sign \\
\colhead{} & \colhead{(MHz)} & \colhead{}
}
\startdata
318& 75.505&\\
445& 76.785&\\
699& 79.345&\\
826& 80.625&\\
1207& 84.46&\\
1334& 85.74&\\ \hline
1963& 92.085&6RTR Perth (92.1 MHz)\\
1969& 92.145&6RTR Perth (92.1 MHz)\\
2223& 94.705&\\
2239& 94.865&6ABCFM Geraldton (94.9 MHz)\\
2240& 94.875&6ABCFM Geraldton (94.9 MHz)\\
2241& 94.885&6ABCFM Geraldton (94.9 MHz)\\
2380& 96.285&\\
2437& 96.865&6SBSFM Perth (96.9 MHz)\\
2439& 96.885&6SBSFM Perth (96.9 MHz)\\
2442& 96.915&6SBSFM Perth (96.9 MHz)\\
2477& 97.265&\\
2557& 98.075&6BAY Geraldton (98.1 MHz)\\
2559& 98.095&6BAY Geraldton (98.1 MHz)\\
2604& 98.545&\\
2637& 98.875&6JJJ Geraldton (98.9 MHz)\\
2638& 98.885&6JJJ Geraldton (98.9 MHz)\\
2639& 98.895&6JJJ Geraldton (98.9 MHz)\\
2717& 99.685&\\
2876& 101.285&\\
\enddata
\tablenotetext{a}{Horizontal line indicates the lower edge of the FM band.}
\end{deluxetable}

Upon individual inspection of the 141 candidate detections, all signals showed very clear signatures of terrestrial RFI in the relevant frequency channel, i.e. strong temporal variability in both the on and off positions, usually persistent across all time steps but occasionally significantly stronger at a single time step.  Thus, all candidate signals at the on position identified from individual dynamic spectrum pixels (0.5\,s and 10\,kHz) can be ruled out via reference to the off signal at the same frequency, at five times the RMS value (250\,Jy/beam). This corresponds to an equivalent isotropic radiated power (EIRP) upper limit of approximately 7\,kW at the distance of \target\ (Table 1).  In comparison, the off position yielded 156 candidate detections above five times the RMS level at that position, comparable to the results from the on position.  

A total of 25 frequency channels out of 2280 useable channels at the on position represents approximately 1\% of the useable channels across the band, consistent with the results of \citet{Offringa:2015}, who found (between 73 and 101\,MHz) that the global occupancy of RFI between 72 and 230\,MHz at the MWA site is 1.13\%.  

In Table~\ref{tab:chans} we identify the likely sources of interference, in the form of the call signs for the FM stations for the likely interferers; the cases easy to identify are due to relatively powerful FM transmitters based in Geraldton (approximately 350\,km from the MWA) or Perth (approximately 700\,km from the MWA).  In these cases, multiple adjacent channels are affected by single interferers, since the typical FM broadcast bandwidth is approximately 100\,kHz (ten of our 10\,kHz channels).  In other cases, the interference may come from FM stations at greater distances, less powerful local transmitters, or local transmitters with non-favourable directionality.  The sources of interference between 75 and 86 MHz (below the FM band) are unknown, but a number of primary services allocated to these frequencies are listed in the Australian Communications and Media Authority (ACMA) Table of Band Allocations\footnote{https://www.acma.gov.au/theacma/australian-radiofrequency-spectrum-plan-spectrum-planning-acma}.  The transmitters responsible for these signals could potentially be at large distances.

\subsection{Persistent Narrow-band Signals}
\begin{figure}
\plotone{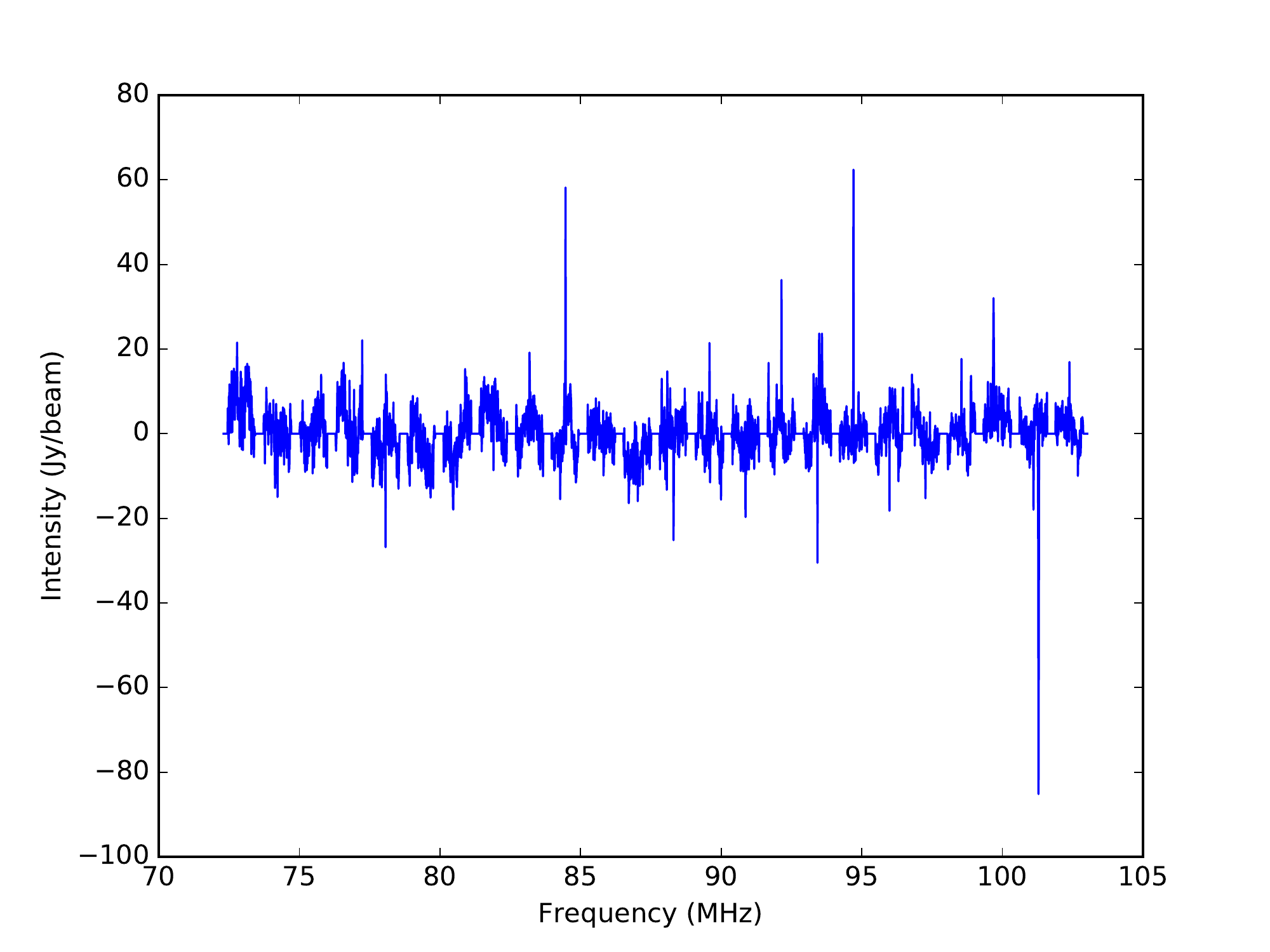}
\caption{Time averaged spectrum at the on position.}
\label{fig:av-spec}
\end{figure}

The time averaged spectrum, over the full duration of the observations, is shown in Figure~\ref{fig:av-spec} for the on position.  The RMS over this spectrum is approximately 6\,Jy/beam.  Adopting a threshold of five times the RMS for candidate detections of persistent narrow-band signals, we find detections above the 30\,Jy/beam threshold at the following centre frequencies (10\, kHz channels): 84.465\,MHz; 92.145\,MHz; 94.705 MHz; 99.685\,MHz; and 101.285\,MHz.  All of these frequencies are accounted for by the frequencies ruled out in the previous section.  Thus, there are no constant narrowband signals from Oumuamua greater than 30\,Jy/beam confined to a 10\,kHz channel, corresponding to an EIRP upper limit of approximately 840\,W.

\subsection{Impulsive Broad-band Signals}
\begin{figure}
\plotone{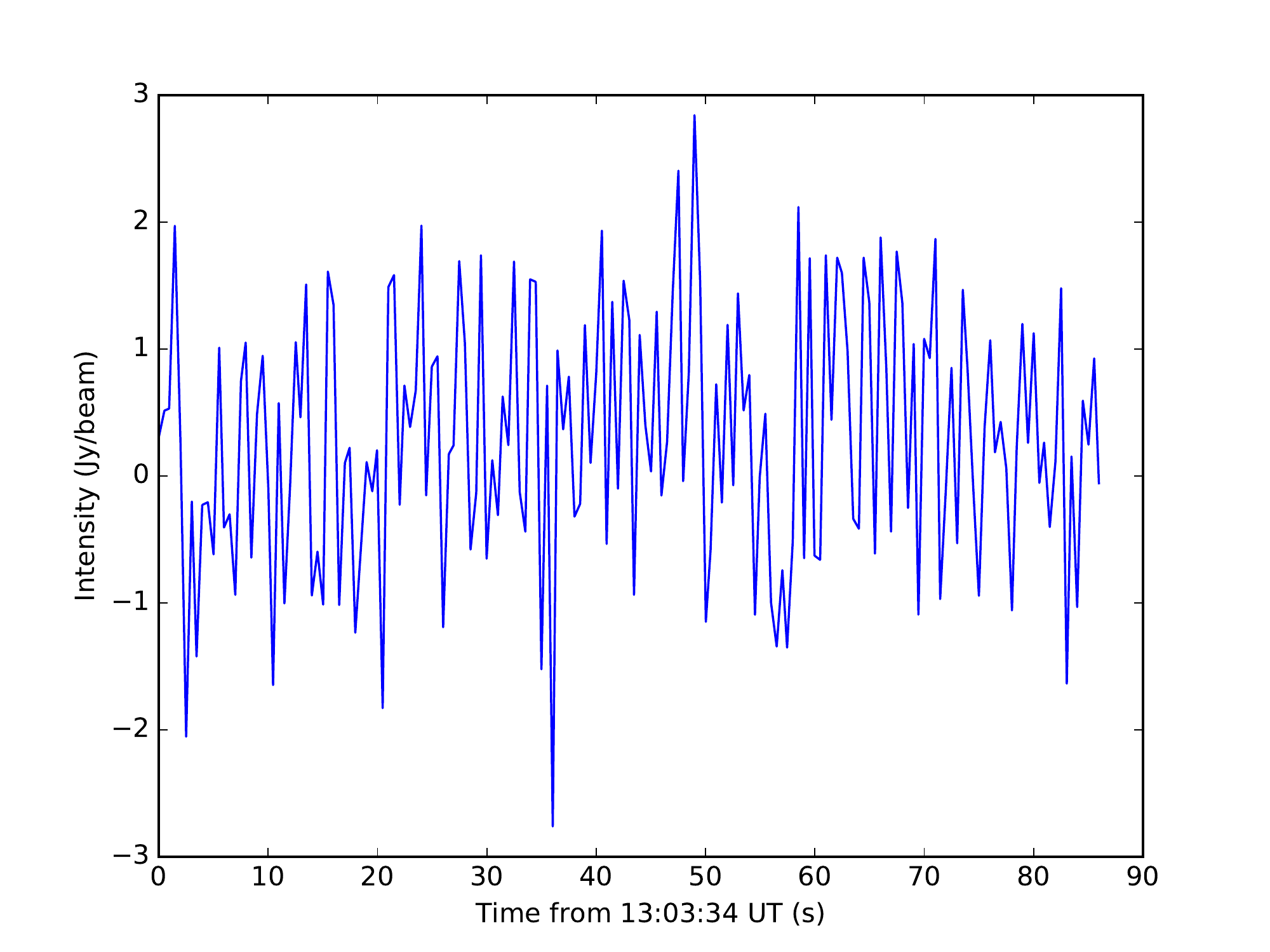}
\caption{Frequency averaged time series at the on position.}
\label{fig:timeseries}
\end{figure}

Finally, the dynamic spectra were averaged over frequency at each 0.5\,s time step (producing a single frequency averaged time series: Figure~\ref{fig:timeseries}) and searched for impulsive, broadband signals.  The RMS of the time series is approximately 1\,Jy/beam and no signals above five times this RMS were detected in the time series.  So, no broadband impulsive signals were detected at 0.5\,s in a 30\,MHz band, corresponding to an EIRP of $\sim100$\,kW.

\section{Discussion and Conclusions}
The limits on EIRP estimated from the data are well within the range of technologies used on Earth by humans.  Therefore it is reasonable to place such transmitter powers at these frequencies within the range of technologies available to other advanced civilisations.  The physical size requirements for such transmitters are also within what is plausible for an object of the size of \target.  

We recall the discussion in the previous section regarding the non-Gaussian statistics of our data and the subjective use of thresholds of five times the measured RMS.  We note that varying these thresholds between three times the RMS and ten times the RMS would result in EIRP limits $\pm$ a factor of two on our estimates.  A factor of two makes no qualitative difference to the conclusions in the previous paragraph.

The higher frequency (1.1--11.6\,GHz) SETI experiment of \citet{2018arXiv180102814E} targeting \target\ also found no signals of interest.  One characteristic of \target\ they accounted for in their observations was the object's rotation period, making sure they sampled all phases of the rotation.  Due to the serendipitous nature of our observations, we were unable to control this coverage (we covered approximately 0.2\% of a rotation period with our observations).  Any loss of sensitivity that would raise our lower limit estimates would depend on the orientation of \target\ and the directionality of the transmitting antenna.  

While our serendipitous observation of \target\ falls partly within the FM broadcast band on Earth, making the discrimination of terrestrial interference from signals originating at \target\ an issue to be overcome, the generally radio quiet location of the MWA (well away from FM and other broadcasters) makes such searches feasible.  For the small number of interfering signals present in our data, the use of on/off positions allowed their identification.  

Moreover, it could be argued that the FM band is a good place to search for transmissions from a spacecraft visiting our Solar System.  It would be quickly apparent to the visitors that the FM band is a highly used frequency range, given the Earth appears to have an approximate 77\,MW EIRP in the FM band \citep{2013AJ....145...23M}, and mainly used for information broadcast rather than two-way communications.  Thus, the FM band would certainly be an interesting frequency range for monitoring and a logical choice for communications --- there would be many people listening on Earth.  However, very few places on Earth are quiet enough in the FM band that such a communication signal could be clearly heard.  The MWA site is one such location.

The MWA is an example of a new generation, flexible system with unique characteristics.  For SETI, a key MWA characteristic is the extreme field-of-view ($>$1000\,deg$^2$) at the frequencies used in this work.  This means that whatever astrophysical target the MWA is looking at, the field-of-view almost certainly contains interesting SETI targets.  The inherent flexibility of the MWA system makes this possible, as long as the data are then appropriately processed for the purposes of SETI.  The MWA and the future SKA will likely have their scientific productivities multiplied due to these inherent possibilities for commensal science.

This paper represents the third opportunistic/serendipitous SETI result with the MWA, after \citet{2016ApJ...827L..22T} and Tingay, Tremblay \& Croft (2018, submitted). In addition to image-based searches such as these, beamforming techniques may be used to generate spectra for individual pixels within the primary beam. By accessing raw voltages from the MWA's Voltage Capture System \citep[VCS;][]{2015PASA...32....5T}, higher spectral resolution than that delivered natively by the correlator can be achieved. Upgrades to the MWA correlator planned for 2018, as well as the planned deployment of additional computing hardware by the Breakthrough Listen project, will enable routine beamforming SETI searches to be performed on any target within the MWA's primary beam, commensally with other science uses of the telescope.

An interesting extension to the imaging mode techniques used in this paper would be to utilise imaging in polarised radio emission.  The great majority of the low frequency radio sky is unpolarised, but intentional transmissions from an object like \target\ may be highly polarised (circularly or linearly), meaning that substantially lower EIRP limits may be possible by searching in polarised emission.  The MWA is proving to be a very capable instrument for measuring polarisation \citep{2018MNRAS.tmp..171O,2017PASA...34...40L} and this mode could be utilised in future SETI experiments.

The fact that commensal opportunities substantially lower the opportunity (and real) costs to perform large SETI experiments, allowing large target lists to be explored more deeply than previously, means that the vanishingly small probability of success for observations of any given SETI target are partially mitigated.  

With all this in mind, although the chance of detecting alien transmissions from \target\ were recognised to be virtually zero, the cost of our experiment was low.  Some predictions say that \target\ is not an unusual object, even though it is the first of its kind to be detected, with the expectation that future facilities may detect large numbers of objects that originate outside the Solar System.  In these conditions, it would be easy to decide to stop subjecting them to SETI experiments, running the risk that the one in a million object is an alien spacecraft.  Lowering the cost of SETI is a key factor in continuing the search.  

The low cost of SETI, combined with the very large sensitivity of the future SKA \citep{2015aska.confE.116S}, provides a compelling future for SETI experiments.

\section*{Acknowledgments}
This scientific work makes use of the Murchison Radio-astronomy Observatory, operated by CSIRO. We acknowledge the Wajarri Yamatji people as the traditional owners of the Observatory site. Support for the operation of the MWA is provided by the Australian Government (NCRIS), under a contract to Curtin University administered by Astronomy Australia Limited. We acknowledge the Pawsey Supercomputing Centre which is supported by the Western Australian and Australian Governments. Funding for Breakthrough Listen research is provided by the Breakthrough Prize Foundation\footnote{\url{http://breakthroughprize.org}}.

\facility{MWA}

\bibliography{biblio} 

\begin{thebibliography}{}
\expandafter\ifx\csname natexlab\endcsname\relax\def\natexlab#1{#1}\fi

\bibitem[{{Altobelli} {et~al.}(2016){Altobelli}, {Postberg}, {Fiege},
  {Trieloff}, {Kimura}, {Sterken}, {Hsu}, {Hillier}, {Khawaja},
  {Moragas-Klostermeyer}, {Blum}, {Burton}, {Srama}, {Kempf}, \&
  {Gruen}}]{2016Sci...352..312A}
{Altobelli}, N., {Postberg}, F., {Fiege}, K., {et~al.} 2016, Science, 352, 312

\bibitem[{{Bracewell}(1960)}]{1960Natur.186..670B}
{Bracewell}, R.~N. 1960, \nat, 186, 670

\bibitem[{{Charnoz} \& {Morbidelli}(2003)}]{2003Icar..166..141C}
{Charnoz}, S., \& {Morbidelli}, A. 2003, \icarus, 166, 141

\bibitem[{Enriquez {et~al.}(2018)Enriquez, Siemion, Lazio, Lebofsky, MacMahon,
  Park, Croft, DeBoer, Gizani, Gajjar, Hellbourg, Isaacson, \&
  Price}]{2018arXiv180102814E}
Enriquez, J.~E., Siemion, A., Lazio, T. J.~W., {et~al.} 2018, Research Notes of
  the AAS, 2, 9

\bibitem[{Fitzsimmons {et~al.}(2017)Fitzsimmons, Snodgrass, Rozitis, Yang,
  Hyland, Seccull, Bannister, Fraser, Jedicke, \&
  Lacerda}]{fitzsimmons2017spectroscopy}
Fitzsimmons, A., Snodgrass, C., Rozitis, B., {et~al.} 2017, Nature Astronomy, 1

\bibitem[{{Freitas}(1980)}]{1980JBIS...33...95F}
{Freitas}, Jr., R.~A. 1980, Journal of the British Interplanetary Society, 33,
  95

\bibitem[{{Isaacson} {et~al.}(2017){Isaacson}, {Siemion}, {Marcy}, {Lebofsky},
  {Price}, {MacMahon}, {Croft}, {DeBoer}, {Hickish}, {Werthimer}, {Sheikh},
  {Hellbourg}, \& {Enriquez}}]{2017PASP..129e4501I}
{Isaacson}, H., {Siemion}, A.~P.~V., {Marcy}, G.~W., {et~al.} 2017, \pasp, 129,
  054501

\bibitem[{{Lenc} {et~al.}(2017){Lenc}, {Anderson}, {Barry}, {Bowman}, {Cairns},
  {Farnes}, {Gaensler}, {Heald}, {Johnston-Hollitt}, {Kaplan}, {Lynch},
  {McCauley}, {Mitchell}, {Morgan}, {Morales}, {Murphy}, {Offringa}, {Ord},
  {Pindor}, {Riseley}, {Sadler}, {Sobey}, {Sokolowski}, {Sullivan},
  {O'Sullivan}, {Sun}, {Tremblay}, {Trott}, \& {Wayth}}]{2017PASA...34...40L}
{Lenc}, E., {Anderson}, C.~S., {Barry}, N., {et~al.} 2017, \pasa, 34, e040

\bibitem[{{McKinley} {et~al.}(2013){McKinley}, {Briggs}, {Kaplan}, {Greenhill},
  {Bernardi}, {Bowman}, {de Oliveira-Costa}, {Tingay}, {Gaensler}, {Oberoi},
  {Johnston-Hollitt}, {Arcus}, {Barnes}, {Bunton}, {Cappallo}, {Corey},
  {Deshpande}, {deSouza}, {Emrich}, {Goeke}, {Hazelton}, {Herne}, {Hewitt},
  {Kasper}, {Kincaid}, {Koenig}, {Kratzenberg}, {Lonsdale}, {Lynch},
  {McWhirter}, {Mitchell}, {Morales}, {Morgan}, {Ord}, {Pathikulangara},
  {Prabu}, {Remillard}, {Rogers}, {Roshi}, {Salah}, {Sault}, {Udaya Shankar},
  {Srivani}, {Stevens}, {Subrahmanyan}, {Wayth}, {Waterson}, {Webster},
  {Whitney}, {Williams}, {Williams}, \& {Wyithe}}]{2013AJ....145...23M}
{McKinley}, B., {Briggs}, F., {Kaplan}, D.~L., {et~al.} 2013, \aj, 145, 23

\bibitem[{Meech {et~al.}(2017)Meech, Weryk, Micheli, Kleyna, Hainaut, Jedicke,
  Wainscoat, Chambers, Keane, Petric, Denneau, Magnier, Berger, Huber,
  Flewelling, Waters, Schunova-Lilly, \& Chastel}]{pmid29160305}
Meech, K., Weryk, R., Micheli, M., {et~al.} 2017, Nature, 552, 378

\bibitem[{{Mitchell} {et~al.}(2008){Mitchell}, {Greenhill}, {Wayth}, {Sault},
  {Lonsdale}, {Cappallo}, {Morales}, \& {Ord}}]{Mitchell:2008}
{Mitchell}, D.~A., {Greenhill}, L.~J., {Wayth}, R.~B., {et~al.} 2008, IEEE
  Journal of Selected Topics in Signal Processing, 2, 707

\bibitem[{{Offringa} {et~al.}(2015){Offringa}, {Wayth}, {Hurley-Walker},
  {Kaplan}, {Barry}, {Beardsley}, {Bell}, {Bernardi}, {Bowman}, {Briggs},
  {Callingham}, {Cappallo}, {Carroll}, {Deshpande}, {Dillon}, {Dwarakanath},
  {Ewall-Wice}, {Feng}, {For}, {Gaensler}, {Greenhill}, {Hancock}, {Hazelton},
  {Hewitt}, {Hindson}, {Jacobs}, {Johnston-Hollitt}, {Kapi{\'n}ska}, {Kim},
  {Kittiwisit}, {Lenc}, {Line}, {Loeb}, {Lonsdale}, {McKinley}, {McWhirter},
  {Mitchell}, {Morales}, {Morgan}, {Morgan}, {Neben}, {Oberoi}, {Ord}, {Paul},
  {Pindor}, {Pober}, {Prabu}, {Procopio}, {Riding}, {Udaya Shankar}, {Sethi},
  {Srivani}, {Staveley-Smith}, {Subrahmanyan}, {Sullivan}, {Tegmark},
  {Thyagarajan}, {Tingay}, {Trott}, {Webster}, {Williams}, {Williams}, {Wu},
  {Wyithe}, \& {Zheng}}]{Offringa:2015}
{Offringa}, A.~R., {Wayth}, R.~B., {Hurley-Walker}, N., {et~al.} 2015, \pasa,
  32, e008

\bibitem[{{Ord} {et~al.}(2010){Ord}, {Mitchell}, {Wayth}, {Greenhill},
  {Bernardi}, {Gleadow}, {Edgar}, {Clark}, {Allen}, {Arcus}, {Benkevitch},
  {Bowman}, {Briggs}, {Bunton}, {Burns}, {Cappallo}, {Coles}, {Corey},
  {deSouza}, {Doeleman}, {Derome}, {Deshpande}, {Emrich}, {Goeke},
  {Gopalakrishna}, {Herne}, {Hewitt}, {Kamini}, {Kaplan}, {Kasper}, {Kincaid},
  {Kocz}, {Kowald}, {Kratzenberg}, {Kumar}, {Lonsdale}, {Lynch}, {McWhirter},
  {Madhavi}, {Matejek}, {Morales}, {Morgan}, {Oberoi}, {Pathikulangara},
  {Prabu}, {Rogers}, {Roshi}, {Salah}, {Schinkel}, {Udaya Shankar}, {Srivani},
  {Stevens}, {Tingay}, {Vaccarella}, {Waterson}, {Webster}, {Whitney},
  {Williams}, \& {Williams}}]{Ord:2010}
{Ord}, S.~M., {Mitchell}, D.~A., {Wayth}, R.~B., {et~al.} 2010, \pasp, 122,
  1353

\bibitem[{{Ord} {et~al.}(2015){Ord}, {Crosse}, {Emrich}, {Pallot}, {Wayth},
  {Clark}, {Tremblay}, {Arcus}, {Barnes}, {Bell}, {Bernardi}, {Bhat}, {Bowman},
  {Briggs}, {Bunton}, {Cappallo}, {Corey}, {Deshpande}, {deSouza},
  {Ewell-Wice}, {Feng}, {Goeke}, {Greenhill}, {Hazelton}, {Herne}, {Hewitt},
  {Hindson}, {Hurley-Walker}, {Jacobs}, {Johnston-Hollitt}, {Kaplan}, {Kasper},
  {Kincaid}, {Koenig}, {Kratzenberg}, {Kudryavtseva}, {Lenc}, {Lonsdale},
  {Lynch}, {McKinley}, {McWhirter}, {Mitchell}, {Morales}, {Morgan}, {Oberoi},
  {Offringa}, {Pathikulangara}, {Pindor}, {Prabu}, {Procopio}, {Remillard},
  {Riding}, {Rogers}, {Roshi}, {Salah}, {Sault}, {Udaya Shankar}, {Srivani},
  {Stevens}, {Subrahmanyan}, {Tingay}, {Waterson}, {Webster}, {Whitney},
  {Williams}, {Williams}, \& {Wyithe}}]{Ord:2015}
{Ord}, S.~M., {Crosse}, B., {Emrich}, D., {et~al.} 2015, \pasa, 32, e006

\bibitem[{{O'Sullivan} {et~al.}(2018){O'Sullivan}, {Lenc}, {Anderson},
  {Gaensler}, \& {Murphy}}]{2018MNRAS.tmp..171O}
{O'Sullivan}, S.~P., {Lenc}, E., {Anderson}, C.~S., {Gaensler}, B.~M., \&
  {Murphy}, T. 2018, \mnras, arXiv:1801.02452

\bibitem[{{Siemion} {et~al.}(2015){Siemion}, {Benford}, {Cheng-Jin},
  {Chennamangalam}, {Cordes}, {Falcke}, {Garrington}, {Garrett}, {Gurvits},
  {Hoare}, {Korpela}, {Lazio}, {Messerschmitt}, {Morrison}, {O'Brien},
  {Paragi}, {Penny}, {Spitler}, {Tarter}, \& {Werthimer}}]{2015aska.confE.116S}
{Siemion}, A., {Benford}, J., {Cheng-Jin}, J., {et~al.} 2015, Advancing
  Astrophysics with the Square Kilometre Array (AASKA14), 116

\bibitem[{{Tingay} {et~al.}(2016){Tingay}, {Tremblay}, {Walsh}, \&
  {Urquhart}}]{2016ApJ...827L..22T}
{Tingay}, S.~J., {Tremblay}, C., {Walsh}, A., \& {Urquhart}, R. 2016, \apjl,
  827, L22

\bibitem[{{Tingay} {et~al.}(2013){Tingay}, {Goeke}, {Bowman}, {Emrich}, {Ord},
  {Mitchell}, {Morales}, {Booler}, {Crosse}, {Wayth}, {Lonsdale}, {Tremblay},
  {Pallot}, {Colegate}, {Wicenec}, {Kudryavtseva}, {Arcus}, {Barnes},
  {Bernardi}, {Briggs}, {Burns}, {Bunton}, {Cappallo}, {Corey}, {Deshpande},
  {Desouza}, {Gaensler}, {Greenhill}, {Hall}, {Hazelton}, {Herne}, {Hewitt},
  {Johnston-Hollitt}, {Kaplan}, {Kasper}, {Kincaid}, {Koenig}, {Kratzenberg},
  {Lynch}, {Mckinley}, {Mcwhirter}, {Morgan}, {Oberoi}, {Pathikulangara},
  {Prabu}, {Remillard}, {Rogers}, {Roshi}, {Salah}, {Sault}, {Udaya-Shankar},
  {Schlagenhaufer}, {Srivani}, {Stevens}, {Subrahmanyan}, {Waterson},
  {Webster}, {Whitney}, {Williams}, {Williams}, \&
  {Wyithe}}]{2013PASA...30....7T}
{Tingay}, S.~J., {Goeke}, R., {Bowman}, J.~D., {et~al.} 2013, \pasa, 30, e007

\bibitem[{{Tremblay} {et~al.}(2015){Tremblay}, {Ord}, {Bhat}, {Tingay},
  {Crosse}, {Pallot}, {Oronsaye}, {Bernardi}, {Bowman}, {Briggs}, {Cappallo},
  {Corey}, {Deshpande}, {Emrich}, {Goeke}, {Greenhill}, {Hazelton},
  {Johnston-Hollitt}, {Kaplan}, {Kasper}, {Kratzenberg}, {Lonsdale}, {Lynch},
  {McWhirter}, {Mitchell}, {Morales}, {Morgan}, {Oberoi}, {Prabu}, {Rogers},
  {Roshi}, {Udaya Shankar}, {Srivani}, {Subrahmanyan}, {Waterson}, {Wayth},
  {Webster}, {Whitney}, {Williams}, \& {Williams}}]{2015PASA...32....5T}
{Tremblay}, S.~E., {Ord}, S.~M., {Bhat}, N.~D.~R., {et~al.} 2015, \pasa, 32,
  e005

\end{thebibliography}

\end{document}